\documentclass[twoside]{article}
\usepackage{fleqn,espcrc2}

\usepackage{graphicx}
\usepackage[figuresright]{rotating}

\newcommand{\AmS}{{\protect\the\textfont2
  A\kern-.1667em\lower.5ex\hbox{M}\kern-.125emS}}

\title{Topological and confining properties of 
 Abelian-projected SU(3)-QCD}

\author{D. Antonov\address{INFN-Sezione di Pisa, 
 Universit\'a degli studi di Pisa, \\
 Dipartimento di Fisica, 
 Via Buonarroti, 2 - Ed. B - I-56127 Pisa, Italy}}

\begin{document}

\begin{abstract}
In this talk, we discuss several topics related to the
Abelian-projected
$SU(3)$-QCD. First of them is the Aharonov-Bohm effect emerging
during the extension of this theory by the 
introduction of the $\Theta$-term. Another topic is devoted to 
various consequences of screening of the dual 
vector bosons by electric vortex loops. In particular, 
it is demonstrated that this effect modifies significantly   
the interaction of quarks. Next, the influence of screening to electric and 
magnetic field correlators in the four-dimensional Abelian-projected
$SU(3)$-QCD is studied. Finally, the bilocal correlator of 
electric field strengths in the three-dimensional gas of $SU(3)$ 
Abelian-projected monopoles is discussed. 
\vspace{1pc}
\end{abstract}

\maketitle

\section{INTRODUCTION}

In the present talk, we shall mostly discuss various nonperturbative 
properties of the effective low-energy theory 
of the $SU(3)$-QCD~\cite{1}, which models confinement of quarks
as the dual Meissner effect~\cite{2}. The partition function 
of this $[U(1)]^2$ magnetically gauge-invariant theory reads 

$$
{\cal Z}=\int \left|\Phi_a\right|{\cal D}\left|\Phi_a\right|
{\cal D}\theta_a{\cal D}{\bf B}_\mu\delta\left(\sum\limits_{a=1}^{3}
\theta_a\right)\times$$

$$\times\exp\left\{-\int d^4x\left[\frac14\left({\bf F}_{\mu\nu}+
{\bf F}_{\mu\nu}^{(c)}\right)^2+\right.\right.$$

$$
+\sum\limits_{a=1}^{3}\left(\frac12\left|\left(\partial_\mu-2ig_m{\bf e}_a
{\bf
B}_\mu\right)\Phi_a\right|^2+\right.
$$

\begin{equation}
\label{1}
\left.\left.\left.+\lambda\left(\left|\Phi_a\right|^2-
\eta^2\right)^2\right)\right]\right\}.
\end{equation}
Here, $g_m$ is the magnetic coupling constant, related to the QCD 
coupling constant $g$ as $g_m=\frac{4\pi}{g}$, and 
${\bf e}_a$'s are the root vectors of $SU(3)$, whose explicit form is 
${\bf e}_1=(1,0)$, ${\bf e}_2=\left(-\frac12,-\frac{\sqrt{3}}{2}\right)$,
${\bf e}_3=\left(-\frac12,\frac{\sqrt{3}}{2}\right)$.
Next, in Eq.~(\ref{1}), ${\bf F}_{\mu\nu}$
stands for the field strength tensor of the field ${\bf B}_\mu$ dual to 
the field ${\bf A}_\mu\equiv\left(A_\mu^3,A_\mu^8\right)$, and 
$\Phi_a=\left|\Phi_a\right|{\rm e}^{i\theta_a}$ are the dual 
Higgs fields describing the condensation of Cooper pairs
of Abelian-projected monopoles.

Note that the phases $\theta_a$'s are related to each other 
by the constraint $\sum\limits_{a=1}^{3}\theta_a=0$, imposed by the 
respective $\delta$-function on the right-hand side of Eq.~(\ref{1}).
This constraint reflects the dependence of the monopoles of three kinds
of each other. Such a dependence is inspired by the fact that the 
monopole magnetic charges are
distributed over the lattice defined by the root vectors, whose sum
vanishes.

Finally, in Eq.~(\ref{1}) we have introduced the field strength tensor 
${\bf F}_{\mu\nu}^{(c)}$ of an external
quark of the colour $c=R,B,G$ (red, blue, green, respectively), which 
obeys the equation $\partial_\mu\tilde{\bf F}_{\mu\nu}^{(c)}=g
{\bf Q}^{(c)}j_\nu$. Here, $\tilde{\cal O}_{\mu\nu}\equiv\frac12
\varepsilon_{\mu\nu\lambda\rho}{\cal O}_{\lambda\rho}$, $j_\nu(x)
\equiv\oint\limits_{C}^{}dx_\nu(\tau)\delta(x-x(\tau))$, and 
${\bf Q}^{(c)}$'s 
are the charges of a quark of the colour $c$ with respect to the Cartan 
subgroup of $SU(3)$: 
${\bf Q}^{(R)}=\left(\frac12,\frac{1}{2\sqrt{3}}\right)$, 
${\bf Q}^{(B)}=\left(-\frac12,\frac{1}{2\sqrt{3}}\right)$, 
${\bf Q}^{(C)}=\left(0,-\frac{1}{\sqrt{3}}\right)$. 
The present lattice data~\cite{3} indicate that in the regime
of the model~(\ref{1}) corresponding to the real QCD, 
the coupling constant 
$\lambda$ is much larger than one, namely $\lambda\simeq 65$.
This makes
it reasonable to consider this model in the London limit, 
$\lambda\to\infty$. In this limit, the model~(\ref{1}) allows for an 
exact string representation,  
and the resulting string effective
action reads~\cite{4},~\cite{komarov}

$$
S_c=\pi^2\int
d^4x\int d^4yD_m^{(4)}(x-y)\times
$$

\begin{equation}
\label{2}
\times\left[\eta^2\bar\Sigma_{\mu\nu}^a(x)
\bar\Sigma_{\mu\nu}^a(y)+\frac{8}{3g_m^2}j_\mu(x)
j_\mu(y)\right].
\end{equation}
Here, $m=g_H\eta$ is the mass of the dual vector bosons with 
$g_H\equiv\sqrt{6}g_m$ standing for their magnetic 
charge, which they acquire due to the Higgs mechanism. Next, 
$D_m^{(4)}(x)=\frac{m}{4\pi^2|x|}K_1(m|x|)$ is the propagator 
of these bosons, where from now on $K_\nu$'s stand for the modified
Bessel functions.
In Eq.~(\ref{2}), we have also introduced the notation 
$\bar\Sigma_{\mu\nu}^a=\Sigma_{\mu\nu}^a-2s_a^{(c)}
\Sigma_{\mu\nu}$. In this expression, $s_a^{(c)}$'s stand 
for certain numbers equal to $0$ and $\pm 1$, which obey the relation 
${\bf Q}^{(c)}=\frac13s_a^{(c)}{\bf e}_a$, and 
$\Sigma_{\mu\nu}^a(x)\equiv\int\limits_{\Sigma_a}^{}d\sigma_{\mu\nu}
(x_a(\xi))\delta(x-x_a(\xi))$ is the vorticity tensor 
current defined on the closed dual string world sheet $\Sigma_a$ 
with $\xi\equiv\left(\xi^1,\xi^2\right)$. 
Note that owing to the one-to-one correspondence existing between 
$\Sigma_{\mu\nu}^a$'s and the multivalued parts of $\theta_a$'s, 
the three vorticity tensor currents are subject to the constraint 
$\sum\limits_{a=1}^{3}\Sigma_{\mu\nu}^a=0$, which stems from the 
analogous constraint imposed on $\theta_a$'s. 
Finally, in the definition of
$\bar\Sigma_{\mu\nu}^a$, we have denoted by $\Sigma_{\mu\nu}$ 
the vorticity tensor current defined on an arbitrary surface 
$\Sigma$ bounded by the contour $C$, which is the world sheet 
of the open dual string, ending up at a quark and an antiquark.

The $\Sigma_{\mu\nu}\times\Sigma_{\mu\nu}$-interaction in the
action~(\ref{2}) can be shown~\cite{5} to describe confinement of
quarks, whereas the $j_\mu\times j_\mu$-interaction 
clearly describes their Yukawa interaction at small distances.
Notice also that in what follows we shall be 
interested in the string effective actions, rather than 
the measure of integration over world-sheet coordinates 
$x_a(\xi)$'s. The Jacobian emerging 
during the change of integration variables $\theta_a\to x_a(\xi)$, 
which should be accounted for in this measure, has been evaluated 
in Ref.~\cite{prd}.

\section{INCLUDING THE $\Theta$-TERM}

Let us now add to the Lagrangian of the model~(\ref{1}) the following 
$\Theta$-term: 

$$\Delta{\cal L}=-\frac{i\Theta g_m^2}{4\pi^2}
\left({\bf F}_{\mu\nu}+{\bf F}_{\mu\nu}^{(c)}\right)\left(\tilde
{\bf F}_{\mu\nu}+\tilde {\bf F}_{\mu\nu}^{(c)}\right).$$ 
In the London limit, 
the string representation of such an extended partition function 
then reads~\cite{6}

$${\cal Z}_{\Theta}^c=\exp\left[-\frac23\left(\frac{(2\pi)^2}{g_m^2}+
\frac{(\Theta g_m)^2}{\pi^2}\right)\times\right.$$

$$\times\oint\limits_{C}^{}dx_\mu
\oint\limits_{C}^{}dy_\mu D_m^{(4)}(x-y)-8(\pi\eta)^2\times$$

$$\left.\times\int d^4x\int d^4y\Sigma_{\mu\nu}(x)D_m^{(4)}(x-y)
\Sigma_{\mu\nu}(y)\right]\times$$

$$\times\left<\exp\left\{2s_a^{(c)}\left[\int d^4x\int d^4y\Biggl(
\frac{i\Theta}{3}\tilde{\bar\Sigma}_{\mu\nu}^a(x)j_\nu(y)\times
\right.\right.\right.$$

$$
\times\partial_\mu D_m^{(4)}(x-y)+2(\pi\eta)^2\Sigma_{\mu\nu}^a(x)
D_m^{(4)}(x-y)\times
$$

\begin{equation}
\label{t1}
\left.\left.\left.
\times\Sigma_{\mu\nu}(y)\Biggr)-\frac{i\Theta}{3}
\hat L(\Sigma_a, C)\right]\right\}\right>_{\Sigma_a}.
\end{equation}
Here, 

$$\hat L\left(\Sigma_a, C\right)\equiv$$

$$\equiv\int d^4x\int d^4y
\tilde\Sigma_{\mu\nu}^a(x)j_\nu(y)\partial_\mu D_0^{(4)}(x-y)$$
is the 4D Gauss linking number of the contour $C$ with the closed 
world sheet $\Sigma_a$, where $D_0^{(4)}(x)=\frac{1}{4\pi^2x^2}$.
The average $\left<\ldots\right>_{\Sigma_a}$
is formally defined with respect to $S_c[\Sigma=0]$, but 
its exact meaning will be discussed below.

The first argument in the first 
exponential factor on the right-hand side of Eq.~(\ref{t1}) 
shows that due to the $\Theta$-term quarks acquire a nonvanishing 
magnetic charge~\cite{wit}, i.e. become dyons. 
We also see that there appear short- and long-ranged interactions 
of such dyons with closed dual strings, as 
well as the short-ranged dyon--open-string interaction.
The above-mentioned 
long-ranged interaction, described by the linking number, 
can be viewed as a scattering of
dyons by the closed dual strings, 
which thus play the r\^ole of solenoids carrying electric
flux. According to Eq.~(\ref{t1}), such a 
scattering, which is nothing else but the four-dimensional 
analogue of the
Aharonov-Bohm effect~\cite{aha}, 
takes place at $\Theta\ne 3\pi\times{\,}({\rm integer})$.
Note that in the $SU(2)$-case, the respective critical values 
of $\Theta$ have been found~\cite{emil} to be equal to 
$2\pi\times{\,}({\rm integer})$.

Let us now carry out the average $\left<\ldots\right>_{\Sigma_a}$, 
taking into account that at zero temperature closed dual strings 
with opposite winding numbers are known to form virtual bound states,
called vortex loops~\cite{popov}. The summation over the grand 
canonical ensemble of these objects has been performed in
Ref.~\cite{loops} and 
yields an effective sine-Gordon theory of two antisymmetric spin-1
tensor fields. In the dilute gas
approximation, which is relevant to the reality since vortex loops 
are only virtual (and therefore small-sized) objects, such a theory 
enables one to calculate correlators of loops. Averaging the 
$\Sigma_a$-dependent exponential factor on the right-hand side of 
Eq.~(\ref{t1}) in the sense of this effective theory by making use of the
cumulant expansion and accounting in this average 
only for the contribution of the dominant, bilocal, irreducible correlator 
of vortex loops, we get 
 
$${\cal Z}_{\Theta}^c=\exp\Biggl\{-8\int d^4x\int d^4y\Biggl[
(\pi\eta)^2\Sigma_{\mu\nu}(x)\times$$

$$\times D_{M_1}^{(4)}(x-y)\Sigma_{\mu\nu}(y)+\frac{i\Theta}{3}
\tilde\Sigma_{\mu\nu}(x)j_\nu(y)\times$$

$$\times\partial_\mu
D_m^{(4)}(x-y)\Biggr]-
\frac23\left(\frac{(2\pi)^2}{g_m^2}+\frac{(\Theta g_m)^2}{\pi^2}\right)
\times$$

$$\times\frac{1}{g_H^2+g_D^2}
\oint\limits_{C}^{}dx_\mu\oint\limits_{C}^{}dy_\mu\times$$

\begin{equation}
\label{t2}
\times\left(g_H^2
D_{M_1}^{(4)}(x-y)+g_D^2D_0^{(4)}(x-y)\right)\Biggr\}.
\end{equation}
Here, $g_D=\frac{2\pi\sqrt{\zeta}}{\Lambda^2}$ describes  
the contribution to the magnetic charge of the dual vector bosons, 
stemming from the Debye screening of those in the gas of electric 
vortex loops. In this formula, 
$\Lambda=\sqrt{\frac{L}{a^3}}$ is the ultraviolet momentum 
cutoff with $L$ standing for the typical distance between the 
neighbours in the gas of vortex loops and $a$ denoting the typical size 
of the loop, $a\ll L$. Next,   
$\zeta\propto {\rm e}^{-S_0}$ is the Boltzmann factor of a 
single vortex loop with the action $S_0$ equal to the string 
tension times the characteristic area of the loop. 

Due to the Debye screening, the 
mass $m$ of the dual vector bosons increases. Since owing to the constraint 
$\sum\limits_{a=1}^{3}\Sigma_{\mu\nu}^a=0$ there exists the 
second independent type of vortex loops, there respectively appears also
the second value of the Debye charge, equal to $g_D\sqrt{3}$. The 
two full masses then read $M_1=\eta\sqrt{g_H^2+g_D^2}$ and  
$M_2=\eta\sqrt{g_H^2+3g_D^2}$.

As it follows from Eq.~(\ref{t2}), 
the above-mentioned screening 
changes drastically the interaction of quarks (dyons). Namely, 
besides the modification of the massive propagator due to the 
additional contribution to the mass of the dual vector bosons, there 
also appears a novel massless interaction. It is also worth noting that, 
as it should be, when the effect of screening is disregarded, 
i.e. $g_D\ll g_H$, one recovers the classical result, which is 
nothing else, but the $\Sigma_a$-independent part of Eq.~(\ref{t1}).

\section{APPLICATIONS TO THE STOCHASTIC VACUUM MODEL}
  
Let us now discuss in more details the confining properties of the
model~(\ref{1}) in the London limit. Namely, let us consider the 
bilocal correlator of electric field strengths, ${\bf f}_{\mu\nu}=
\partial_\mu{\bf A}_\nu-\partial_\nu{\bf A}_\mu$, 
which plays the 
major r\^ole in the so-called stochastic vacuum model~\cite{svm} 
(see Ref.~\cite{rev1} for reviews).  
Within this model, such a correlator is parametrized by the two 
coefficient functions as follows:

$$\left<f_{\mu\nu}^i(x)f_{\lambda\rho}^j(0)\right>_{{\bf A}_\mu, 
{\bf j}_\mu^{\rm m}}=\delta^{ij}\left\{\left(\delta_{\mu\lambda}
\delta_{\nu\rho}-\delta_{\mu\rho}\delta_{\nu\lambda}\right)\times
\right.$$

$$\times D\left(x^2
\right)+\frac12\left[\partial_\mu\left(x_\lambda\delta_{\nu\rho}-
x_\rho\delta_{\nu\lambda}\right)+\right.$$

\begin{equation}
\label{correl}
\left.\left.+\partial_\nu\left(
x_\rho\delta_{\mu\lambda}-x_\lambda\delta_{\mu\rho}\right)\right]
D_1\left(x^2\right)\right\}.
\end{equation}
In this formula, $\left<\ldots\right>_{{\bf A}_\mu}$ stands for the
average over 
free diagonal gluons, and $\left<\ldots\right>_{{\bf j}_\mu^{\rm m}}$
denotes a certain average over monopoles, which provides the  
condensation of their Cooper pairs. It is the coupling of the 
${\bf B}_\mu$-field, dual to the field ${\bf A}_\mu$, to 
${\bf j}_\mu^{\rm m}$'s,
which makes both functions $D$ and $D_1$ 
nontrivial and finally adequate to those of the real QCD.
Namely, these functions turn out to have the following form~\cite{JH}:

$$
D=\frac{m^2M_1}{4\pi^2}\frac{K_1(M_1|x|)}{|x|},
$$

$$
D_1=\frac{g_D^2}{\pi^2\left(g_H^2+g_D^2\right)|x|^4}+
\frac{m^2}{2\pi^2M_1x^2}\times$$

$$\times\left[\frac{K_1(M_1|x|)}{|x|}+\frac{M_1}{2}
\left(K_0(M_1|x|)+K_2(M_1|x|)\right)\right].$$
In the infrared limit, $|x|\gg M_1^{-1}$, the respective asymptotic
behaviours read

\begin{equation}
\label{as1}
D\to\frac{(mM_1)^2}{4\sqrt{2}\pi^{\frac32}}
\frac{{\rm e}^{-M_1|x|}}{(M_1|x|)^{\frac32}},
\end{equation}

\begin{equation}
\label{as2}
D_1\to\frac{g_D^2}{\pi^2\left(g_H^2+g_D^2\right)|x|^4}+
\frac{(mM_1)^2}{2\sqrt{2}\pi^{\frac32}}
\frac{{\rm e}^{-M_1|x|}}{(M_1|x|)^{\frac52}}.
\end{equation}
One can see that according to the above-presented asymptotics, the 
inverse correlation length of the vacuum in the 
model under study is equal to the  
smallest of the two full masses, $M_1$. Besides that, owing to the
screening, the function $D_1$ contains the novel nonperturbative 
$\frac{1}{|x|^4}$-term, which might be important for modelling the 
L\"uscher term~\cite{lu} in the quark-antiquark potential~\cite{dosch}. 
Subtracting from the function $D_1$ this contribution, which has the 
same functional form as the pure perturbative contribution in the 
real QCD, we see that the remained 
part of the asymptotics~(\ref{as2}) together with the
asymptotics~(\ref{as1}) 
are in a good agreement with the QCD lattice measurements 
of the functions $D$ and $D_1$~\cite{lattice}  
(see Ref.~\cite{develop} for recent developments and Ref.~\cite{rev2} 
for reviews). Namely, due to the preexponential behaviour, $D\gg D_1$ 
in the infrared limit, $|x|\gg M_1^{-1}$, whereas in the 
ultraviolet limit, $D\ll D_1$. 

It is also worth noting
that when the screening is disregarded, i.e. $g_D\ll g_H$, the 
inverse correlation length of the vacuum goes over to $m$, and the classical 
expressions~\cite{4} (see Ref.~\cite{class} for the $SU(2)$-case) for the
functions $D$ and $D_1$ recover. 
In particular, the nonperturbative screening-motivated  
$\frac{1}{|x|^4}$-contribution 
to the function $D_1$ vanishes in this limit.

Besides the above-discussed modifications of the classical expressions 
for the correlators of electric field strengths inspired by screening, 
this effect changes also the classical 
expression for the propagator of the 
dual vector bosons. Indeed, the latter one reads 

\begin{equation}
\label{BB}
\left<B_\mu^a(x)B_\nu^b(0)\right>=\delta^{ab}\delta_{\mu\nu}
D_m^{(4)}(x)
\end{equation}
with $B_\mu^a\equiv {\bf e}_a{\bf B}_\mu$, whereas with the account
for screening it changes to~\cite{JH}

$$\delta_{\mu\nu}\left\{\frac13D_m^{(4)}(x)+\frac{g_D^2}{2}\left[
2\frac{g_H^2+2g_D^2}{\left(g_H^2+3g_D^2\right)\left(g_H^2+g_D^2\right)}
\times\right.\right.$$

$$
\times 
D_0^{(4)}(x)+\frac{g_H^2}{g_D^2}\left(\frac{1}{3\left(g_H^2+3g_D^2
\right)}D_{M_2}^{(4)}(x)+\right.
$$

\begin{equation}
\label{diagon}
\left.\left.\left.+\frac{1}{g_H^2+g_D^2}
D_{M_1}^{(4)}(x)\right)\right]\right\}
\end{equation}
for $a=b$ and  

$$\delta_{\mu\nu}\left\{\frac13D_m^{(4)}(x)+\frac{g_D^2}{2}\left[
-\frac{2g_D^2}{\left(g_H^2+3g_D^2\right)\left(g_H^2+g_D^2\right)}
\times\right.\right.$$

$$\times D_0^{(4)}(x)+\frac{g_H^2}{g_D^2}\left(
\frac{1}{3\left(g_H^2+3g_D^2\right)}D_{M_2}^{(4)}(x)-\right.$$

\begin{equation}
\label{off}
\left.\left.\left.-\frac{1}{g_H^2+g_D^2}D_{M_1}^{(4)}(x)
\right)\right]\right\}
\end{equation}
for $a\ne b$. We see that the screening makes the propagator 
of the dual vector bosons nonvanishing even for $a\ne b$. 
As it should be, in the limit when the screening is disregarded, 
$g_D\ll g_H$, these off-diagonal components of the propagator 
given by Eq.~(\ref{off}) vanish, whereas the diagonal ones given 
by Eq.~(\ref{diagon}) go over to the classical expression, so that 
Eq.~(\ref{BB}) recovers.

Contrary to the four-dimensional case, in three dimensions 
the present lattice data allow one 
to assume that Abelian-projected monopoles 
form a gas. Such a gas of $SU(3)$-monopoles has for the
first time been considered 
in Ref.~\cite{su3}, and its partition function reads

$${\cal Z}=1+\sum\limits_{N=1}^{\infty}
\frac{\zeta^N}{N!}\left(\prod\limits_{n=1}^{N}
\int d^3z_n\sum\limits_{a_n=\pm 1,\pm 2, \pm 3}^{}
\right)\times$$

\begin{equation}
\label{mongas}
\times\exp\left[-\frac{g_m^2}{4\pi}\sum\limits_{n<k}^{}
\frac{{\bf e}_{a_n}{\bf e}_{a_k}}{|{\bf z}_n-{\bf z}_k|}
\right].
\end{equation}
In this formula, $\zeta\propto\exp\left(-\frac{{\rm const}}{g^2}
\right)$ stands for the Boltzmann factor of a single monopole, and 
${\bf e}_{-a}=-{\bf e}_a$. The 
string representation of the 
Wilson loop in this gas, constructed in Ref.~\cite{epl}, turned out  
to be alternative to the one of the $SU(2)$-case, found  
in Ref.~\cite{su2}. (See Ref.~\cite{nikita} for the discussion 
of Polyakov loops and their correlators in the $SU(2)$ monopole gas.) 
By virtue of this representation in the approximation when the 
monopole gas is so dilute that its density is much less than $\zeta$, 
one can deduce the respective expressions 
for the functions $D$ and $D_1$. In the model under study, those 
are defined by Eq.~(\ref{correl}) with the average $\left<\ldots
\right>_{{\bf j}_\mu^{\rm m}}$ replaced by the average with respect 
to the partition function~(\ref{mongas}). Together with the contribution 
of the free diagonal gluons to the function $D_1$, these two functions
read 

\begin{equation}
\label{d}
D=12\pi\zeta\frac{{\rm e}^{-m|{\bf x}|}}{|{\bf x}|},
\end{equation}

\begin{equation}
\label{d1}
D_1=\frac{24\pi\zeta}{(m|{\bf x}|)^2}\left(m+\frac{1}{|{\bf x}|}
\right){\rm e}^{-m|{\bf x}|}.
\end{equation}
Here, $m=g_m\sqrt{3\zeta}$ is the Debye mass of the two 
scalar bosons, dual to the diagonal gluons. Similarly to the 
above-considered four-dimensional case, we see that 
Eqs.~(\ref{d}) and (\ref{d1}) well agree with the lattice
calculations in the real QCD~\cite{lattice},~\cite{develop},~\cite{rev2}. 
In particular, the inverse correlation length of the vacuum is now equal 
to $m$, and at the distances larger than this length, 
$D\gg D_1$ due to the preexponential behaviour.

\section{CONCLUSIONS}
In the present talk, we have discussed various nonperturbative properties 
of Abelian-projected $SU(3)$-QCD in four and three dimensions. In the 
four-dimensional case, we have firstly considered the respective dual
Abelian Higgs type model extended by the introduction of the
$\Theta$-term. In this way, the string representation of
such an extended model has been derived, which has in particular
demonstrated how the $\Theta$-term  
leads to the appearance of the magnetic charge of external quarks,
making out of them dyons. The critical values of $\Theta$, at which 
the Aharonov-Bohm scattering of dyons over the closed dual strings
disappears, have been found. Next, the effect of the Debye screening of
the dual vector bosons by virtual electric vortex loops, built 
out of the closed dual strings, has been taken into account. 

Then, the confining properties of the four- and three-dimensional 
$SU(3)$ Abelian-projected theories within the 
stochastic vacuum model have been addressed. In particular, in the 
four-dimensional case the r\^ole of the above-mentioned screening has been 
discussed. The influence 
of this effect to the propagators of the dual vector bosons has 
also been considered. Finally, the bilocal correlator of electric 
field strengths in the dilute three-dimensional gas of $SU(3)$ 
Abelian-projected monopoles has been evaluated.

In conclusion, the performed investigations have shown that
Abelian-projected theories are not only adequate to the description of 
confinement of quarks in QCD, but possess also a lot of interesting 
nonperturbative properties themselves.

\section{ACKNOWLEDGMENTS}
The author is indebted to Prof. A. Di Giacomo for useful
discussions and hospitality, and INFN for the financial support. 
He is also greatful to the organizers 
of the Euroconference "QCD 00" (Montpellier, France, 6-13th July 2000) 
for an opportunity to present these results in a very stimulating 
atmosphere.

\section{DISCUSSION}
{\bf Dr. N. Brambilla (University of Heidelberg)}: {\it You have
mentioned 
that the London limit is in the agreement with the lattice results, but 
I would like to point out that a lot of the  
most recent lattice data show that 
the QCD vacuum looks like a dual superconductor at the border of 
type-I and type-II. Could you please comment on this issue.}

{\bf D. Antonov}: {\it My statement was based on the lattice
data of Ref.~\cite{3}, which demonstrated that in the QCD-relevant 
regime of the 
effective dual theory considered,  
the coupling constant of the dual Higgs potential 
should be much larger than unity.}

\end{document}